\documentclass[runningheads]{llncs}
\usepackage[T1]{fontenc}
\usepackage{latexsym}
\usepackage{amssymb}
\usepackage{amsmath}
\usepackage{amsthm}
\usepackage{booktabs}
\usepackage{enumitem}
\usepackage{graphicx}
\usepackage{color}
\usepackage{algorithm}
\usepackage[utf8]{inputenc}
\usepackage{algpseudocode}
\usepackage{array}
\usepackage{multirow} 

\DeclareUnicodeCharacter{02BE}{'}
\DeclareUnicodeCharacter{02BF}{`}
\DeclareUnicodeCharacter{1EA1}{\d{a}} % ạ → a with dot below
\DeclareUnicodeCharacter{1ECD}{\d{o}} % ọ → o with dot below
\DeclareUnicodeCharacter{1EE3}{\d{o}}     % ợ ≈ o with dot
\DeclareUnicodeCharacter{1EDF}{\~{o}}     % ỡ ≈ o with tilde
\DeclareUnicodeCharacter{01B0}{u}         % ư  ≈ plain u
\DeclareUnicodeCharacter{1EAD}{\d{\^a}}     % ậ = a + circumflex + dot below
\DeclareUnicodeCharacter{1EA3}{\text{\u a}} % ả = a + hook above → approximated with breve
\DeclareUnicodeCharacter{01B0}{u}           % ư → approximated as plain u (no horn)
\DeclareUnicodeCharacter{1EDF}{\~{o}}    % ỡ ≈ o with tilde (no horn support in pdflatex)
\DeclareUnicodeCharacter{1EA7}{\`{\^a}}  % ầ = a with circumflex and grave

\newcommand{\BibTeX}{B\kern-.05em{\sc i\kern-.025em b}\kern-.08em\TeX}

\title{\vspace{-2em}Strategic Communication and Language Bias in Multi-Agent LLM Coordination}

\author{\vspace{-2em}Anonymous submission}

\author{
Alessio Buscemi\inst{1}\orcidID{0009-0003-4668-9915} \and
Daniele Proverbio\inst{2}\orcidID{0000-0002-0122-479X} \and
Alessandro Di Stefano\inst{3}\orcidID{0000-0003-4905-3309} \and
The Anh Han\inst{3}\orcidID{0000-0002-3095-7714} \and
German Castignani \inst{1}\orcidID{0000-0001-5594-4904} \and
Pietro Liò\inst{4}\orcidID{0000-0002-0540-5053}
}

\institute{1- Luxembourg Institute of Science and Technology, 4362 Esch-Belval, Luxembourg \email{alessio.buscemi@list.lu, german.castignani@list.lu} \\
2- University of Trento, 38123 Trento, Italy \email{daniele.proverbio@unitn.it} \\
3- University of Teesside, TS1 3BX Middlesbrough, UK \email{A.DiStefano@tees.ac.uk, t.han@tees.ac.uk}\\
4- University of Cambridge, CB2 1TN Cambridge, UK \email{pl219@cam.ac.uk}
}

\titlerunning{Communication and Language Bias in Multi-Agent LLMs}
\authorrunning{A. Buscemi et al.}

\begin{document}

%%%%%%%%%%%%%%%%%%%%%%%%%%%%%%%%%%%%%%%%%%%%%%%%%%%%%%%%%%%%%%%%%%%%%%%%
\maketitle

\begin{abstract}

Large Language Model (LLM)-based agents are increasingly deployed in multi-agent scenarios where coordination is crucial but not always assured. Research shows that the way strategic scenarios are framed linguistically can affect cooperation. This paper explores whether allowing agents to communicate amplifies these language-driven effects. Leveraging FAIRGAME \cite{buscemi2025fairgame}, we simulate one-shot and repeated games across different languages and models, both with and without communication. Our experiments, conducted with two advanced LLMs—GPT-4o and Llama 4 Maverick—reveal that communication significantly influences agent behavior, though its impact varies by language, personality, and game structure. These findings underscore the dual role of communication in fostering coordination and reinforcing biases.

\end{abstract}

\section{Introduction}
\label{sec:introduction}

AI agents powered by large language models (LLMs) are increasingly influencing research~\cite{lu2024llms}, social dynamics~\cite{tessler2024ai}, and industrial processes~\cite{patel2020leveraging,stone2020artificial}. While numerous definitions of AI agents exist, this work adopts a minimal operational view: an agent is defined as any entity that makes decisions based on an understanding of its context. Since our focus is on decision-making behavior, we abstract away other capacities (e.g., perception) and study how agents act within simulated environments. Although these decisions have no direct real-world consequences in our experiments, their potential operational significance becomes apparent when such agents are embedded in larger socio-technical systems.

Understanding how LLM-based agents behave and interact in multi-agent settings is essential for both innovation and fairness. However, current efforts remain largely focused on single-agent explainability~\cite{ali2025entropy,el2025towards} and bias detection~\cite{li2023halueval}. Multi-agent environments introduce emergent biases that simple human behavior emulators often fail to anticipate~\cite{hammond2025multi,park2023generative}, and these biases already manifest in domains such as automated dispute resolution, pricing, and supply chain negotiations~\cite{brooks2022artificial,falcao2024making}.
Game theory~\cite{owen2013game} provides a principled framework for studying strategic interactions and has recently been extended to LLM-driven agents~\cite{capraro2024language,fontana2024nicer,wang2024large}. Yet, reproducible methodologies that bridge theory and empirical testing in multi-agent contexts remain limited~\cite{mao2023alympics}. FAIRGAME~\cite{buscemi2025fairgame} is a framework that addresses this gap by simulating game-theoretic scenarios ranging from classical games to complex industrial and governance applications~\cite{balabanova2025media}. This framework enables the configuration of agents with distinct strategic, linguistic, and cultural traits, and logs outcomes at scale~\cite{han2021or,buscemi2025llms,montero2022inferring}. Previous experiments~\cite{buscemi2025fairgame,buscemi2025llms} revealed that the linguistic cues embedded in agents' own prompts influenced cooperation levels, even in the absence of explicit communication, uncovering a hidden channel through which bias shapes collective behavior.

Building on these findings, the present study investigates how explicit communication between agents influences emerging collective behavior. We extend the \textsc{FAIRGAME} framework to support inter-agent dialogue and systematically compare interactive and non-interactive conditions in controlled game-theoretic environments. The aim is to determine whether communication mitigates, amplifies, or keeps latent language biases unchanged, depending on the structure of incentives and prevailing conversational norms. To this end, we conduct a large-scale evaluation across two canonical game settings, using two LLMs, three languages, and varied experimental configurations. We further examine the characteristics of agent communication, such as message length and stylistic features, to understand how strategic intent is conveyed. Finally, we identify directions for future research on the role of communication in shaping cooperative dynamics within multi-agent LLM systems.

\section{Methodology}
\label{sec:methodology}

This section outlines the methodology adopted in this study, detailing the selected framework, the language models used, the languages chosen, and all relevant experimental settings.

\vspace{-2mm}
\subsection{Framework}
\label{sub:framework}

FAIRGAME~\cite{buscemi2025fairgame} was selected for this study as a computational framework to provide a practical, systematic and configurable base to simulate strategic interactions among LLM-based agents. Its design (fully detailed in ~\cite{buscemi2025fairgame}) supports a wide variety of game-theoretic settings through prompt-based interfaces, allowing researchers to define agent traits, payoff structures, and interaction parameters using natural language. This flexibility facilitates controlled experimentation across models, languages, and behavioral setups without requiring substantial modifications to the underlying architecture.
As inputs, FAIRGAME takes (i) a \emph{Configuration File}, which specifies game parameters, payoff matrices, and agent characteristics (e.g., personality traits ~\cite{fan2024comp,he2024afspp,newsham2025inducing}); and (ii) a \emph{Prompt Template}, which defines the textual instructions presented to agents during each round. The template dynamically incorporates relevant configuration values, including game type, strategies, and agent-specific variables. For repeated games, it also integrates a historical context. Templates can be written in any language, enabling multilingual experimentation.

This study extends the investigation by focusing on the impact of communication, which was introduced but not further explored in the original FAIRGAME study~\cite{buscemi2025fairgame}. Before choosing an action in each round, agents exchange one message that is informed by the full game context, including prior exchanges. These messages are visible to all agents and are included in the strategy prompt for the current round. This addition allows to investigate how explicit communication shapes agents' behavior, coordination, and strategic alignment, factors that are not observable in settings based on silent interactions.

\vspace{-2mm}
\subsection{LLMs and languages}
\label{sub:llms}

We evaluate AI agents using two state-of-the-art LLMs, summarized in Table~\ref{tab:model_descriptions} along with their key characteristics. All models were tested under the default settings recommended by their respective providers. For each LLM, we employed the most recent version available at the time of experimentation, conducted between 3 and 23 July 2025.

\vspace{-5mm}
\begin{table}[h!]
\scriptsize
\centering
\makebox[\textwidth]{%
\begin{tabular}{|p{1.2cm}|p{1.3cm}|p{1.3cm}|p{1.7cm}|p{1.5cm}|p{2.3cm}|p{1.9cm}|} 
    \hline
    \textbf{Model} & \textbf{Provider} & \textbf{No. Params.} & \textbf{Licensing Type} & \textbf{Entry Point} &  \textbf{Version} & \textbf{Configuration} \\
    \hline
    Llama 4 Maverick (2025-04-05) & Meta Platforms & 400B & Open-source & Replicate API & meta/llama-4-maverick-instruct & Temperature: 0.6; Top\_p: 1.0\\ 
    \hline
    GPT-4o (2024-11-20) & OpenAI, Inc. & N/A & Proprietary & OpenAI API & gpt-4 & Temperature: 1.0; Top\_p: 1.0 \\
    \hline
\end{tabular}%
}
\vspace{0.5mm}
\caption{Overview of the evaluated LLMs and their configurations.}
\label{tab:model_descriptions}
\end{table}
\vspace{-9mm}
Our evaluation is conducted in three languages: English, Arabic, and Vietnamese, to promote coverage across diverse linguistic and cultural contexts. These languages cover variation in scripts, grammatical structures, and geographical distribution, enabling a broad analysis of potential language biases.

Each game template was initially created in English and subsequently translated into Arabic and Vietnamese using automated translation tools. All translations were then reviewed and refined by native speakers to ensure linguistic accuracy and cultural appropriateness. The personality traits associated with agents were also manually adapted by native speakers in each target language.

\vspace{-2mm}
\subsection{Games}
\label{sub:games}

We examined two canonical game-theoretic scenarios to explore how language and communication shape LLM agent behavior and potential biases. The Prisoner's Dilemma models social dilemmas where agents must choose between mutual cooperation or individual gain through defection. Patterns of low cooperation--especially when identity cues differ-- may reveal emergent bias or distrust. The Battle of Sexes captures coordination under asymmetric preferences: consistent alignment with one agent’s preferred outcome can signal dominance or bias in negotiation. These games jointly illuminate how linguistic framing and communication impact both cooperative dynamics and the reinforcement of social asymmetries.

Both games follow standard formulations~\cite{owen2013game} and are encoded using the FAIRGAME configuration template. Each game is defined through a payoff matrix, grouping penalties or rewards for the two players in a generic structure:
\vspace{-1mm}
\[
\begin{pmatrix}
x_{1,1} = (a_1, a_2) & \quad x_{1,2} = (b_1, b_2) \\
x_{2,1} = (c_1, c_2) & \quad x_{2,2} = (d_1, d_2)
\end{pmatrix}.
\]
To assess strategic sensitivity, we implemented three variants \cite{wang2015universal} of the Prisoner's Dilemma, modifying the \textit{dilemma strength}, i.e. the difference between the mutual defection and mutual cooperation penalties. The \textit{conventional} configuration uses $x_{1,1} = (6,6)$, $x_{1,2} = (0,10)$, $x_{2,1} = (10,0)$, and $x_{2,2} = (2,2)$, yielding a dilemma strength of 4. The \textit{harsh} version sets $x_{1,1} = (8,8)$ and $x_{2,2} = (5,5)$, with a strength of 3. The \textit{mild} version keeps $x_{1,1} = (8,8)$ but sets $x_{2,2} = (2,2)$, increasing the strength to 6.
For the Battle of Sexes, we adopted the standard asymmetric configuration of rewards: $x_{1,1} = (10,7)$, $x_{1,2} = (0,0)$, $x_{2,1} = (0,0)$, and $x_{2,2} = (7,10)$. This setting emphasizes conflicting preferences with a shared incentive to coordinate, making it ideal for analyzing alignment behavior.

\vspace{-2mm}
\subsection{Set up}
\label{sub:set up}

Each experiment consists of either one-shot games of 1 round or repeated games comprising 10 rounds, with no early termination criteria, and is conducted separately for each LLM listed in Table~\ref{tab:model_descriptions}. We test both conditions in which agents are explicitly informed of the total number of rounds and conditions in which this information is omitted, as this may influence strategic behaviour and deviations from game-theoretic predictions~\cite{axelrod1981evolution}.

Prior research \cite{fontana2024nicer,wang2024large} demonstrates that LLMs in their default settings do not always comply with predictions from game theory.
Instead, they exhibited consistently cooperative behavior when engaging in traditional game-theoretic scenarios, due to a likely knowledge bias on the games. 
For this reason, we here evaluate the influence of personality traits, by defining each agent as either \textit{cooperative} or \textit{selfish}, two archetypes widely used and extensively validated in the game theory literature. Agent identifiers are intentionally neutral (\texttt{agent1} and \texttt{agent2}) to avoid introducing biases or unintended priming effects that could compromise the interpretability of the results. 
Personality traits are consistently translated into all tested languages to ensure semantic fidelity, while agent identifiers remain untranslated and serve purely as placeholders. Agents are not made aware of their opponent's personality type. All permutations of personality pairings are systematically evaluated, including homogeneous setups and heterogeneous configurations (one cooperative and one selfish agent).
To assess the role of communication, we replicate each experiment under two conditions: with and without inter-agent communication.

For statistical reliability, every game configuration is executed 10 times. This results in a total of 4,320 individual games for the Prisoner's Dilemma and 1,420 for the Battle of Sexes.
Across all runs, agents make 47,520 individual decisions in the Prisoner's Dilemma and 15,840 in the Battle of Sexes.
In the communication-enabled scenarios, agents exchange a total of 23,760 messages in the Prisoner's Dilemma and 7,920 in the Battle of Sexes.

\section{Results}
\label{sec:results}

%In this section we report and discuss the results obtained in the experiments described in Section \ref{sec:methodology}.

\subsection{Prisoner's Dilemma}
\label{sub:results_prisoner}

\begin{figure*}[t]
    \makebox[\textwidth][c]{%
        \includegraphics[width=1\linewidth]{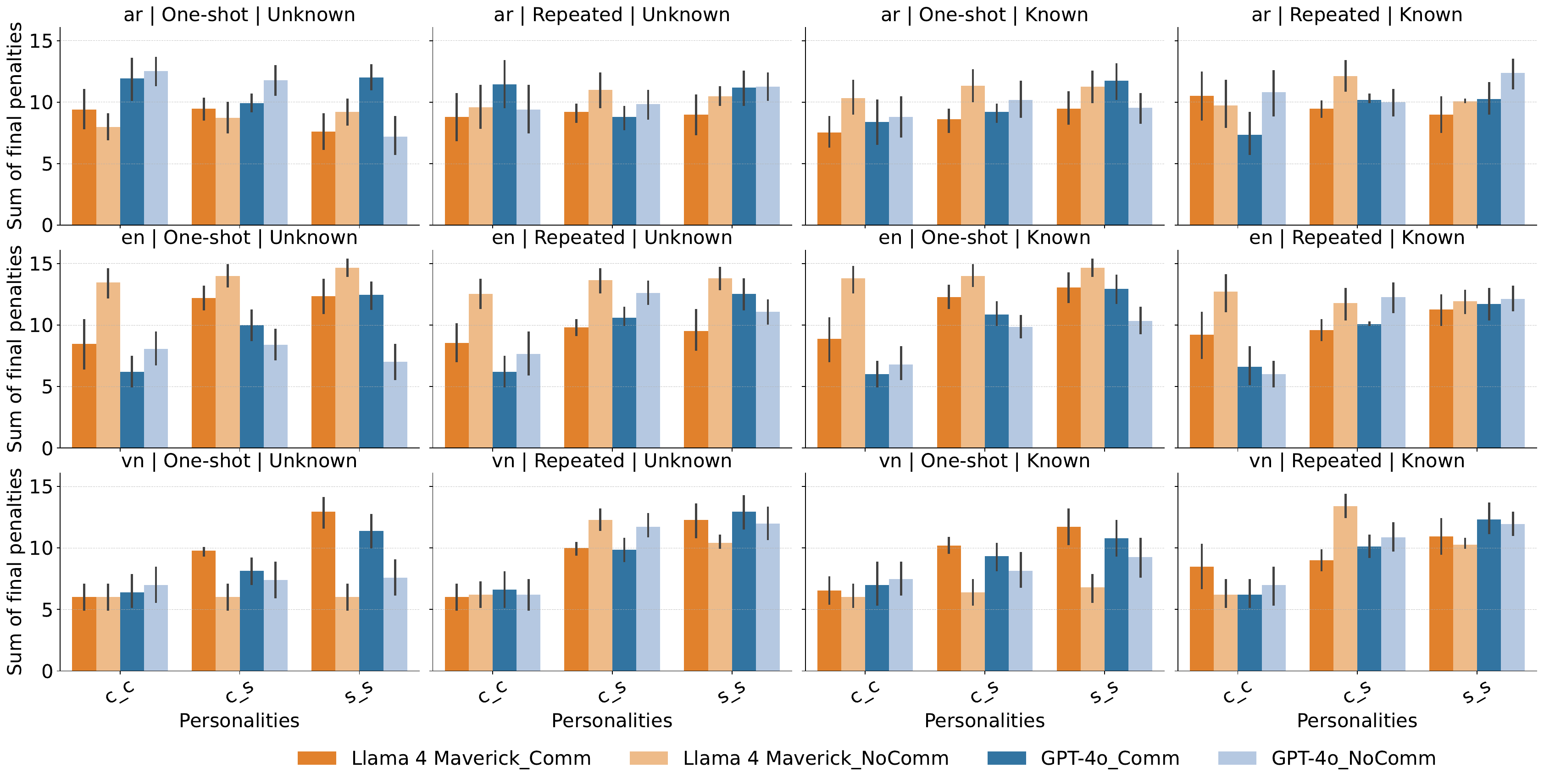}%
    }
     \vspace{-7mm}
    \caption{Impact of communication in the Prisoner's Dilemma game across the considered LLMs. Each bar corresponds to the sum of the penalties obtained by both agents at the end of the game. Each plot corresponds to a combination of language (ar/en/vn) | {type of game (repeated/one-shot) | awareness of the number of rounds (known/unknown). 95\% confidence intervals are reported for each bar.}}
    \label{fig:results_prisoner}
    \vspace{-2mm}
\end{figure*}

Figure~\ref{fig:results_prisoner} shows the final penalties accumulated by LLM agents in the Prisoner's Dilemma. Lower penalties indicate higher levels of cooperation (i.e., the agent consistently chose options that benefit both parties, even at the cost of individual gain). From a practical perspective, high cooperation reflects a greater willingness to treat the other agent fairly. Lower cooperation levels (increased penalties) suggest emerging biases, such as distrust derived from language use.

When communication is conducted in Arabic (top row), Llama 4 Maverick consistently reduces its penalties across both one-shot and repeated settings. A similar trend is observed in most English experiments (middle row), although the magnitude of reduction is higher.
In contrast, communication in Vietnamese (bottom row) leads to an increase in penalties for Llama 4 Maverick (dark orange bars consistently higher than light orange ones). However, it is worth noting that average penalties in Vietnamese are still lower than those recorded in English or Arabic.
For GPT-4o, communication generally results in slightly higher penalties across all three languages, indicating reduced cooperation.
Lastly, the personality pairings alter the patterns: scenarios involving selfish agents tend to yield higher penalties, especially in English and Vietnamese when communication is involved, while penalties tend to be rather constan in Arabic.

Figure 2 shows the evolution of average strategy values over 10 rounds in the repeated Prisoner's Dilemma, across all languages and personality configurations. Llama 4 Maverick-based agents exhibit greater stability when communication is enabled.
When communication is disabled, however, defection tends to rise after the first round. Enabling communication is thus associated with more sustained cooperation throughout the game.
For GPT-4o, however, the level of defection gradually increases over time, and the difference between the communication and no-communication conditions is less pronounced.

\begin{figure*}[t]
    \centering
    \includegraphics[width=1\linewidth]{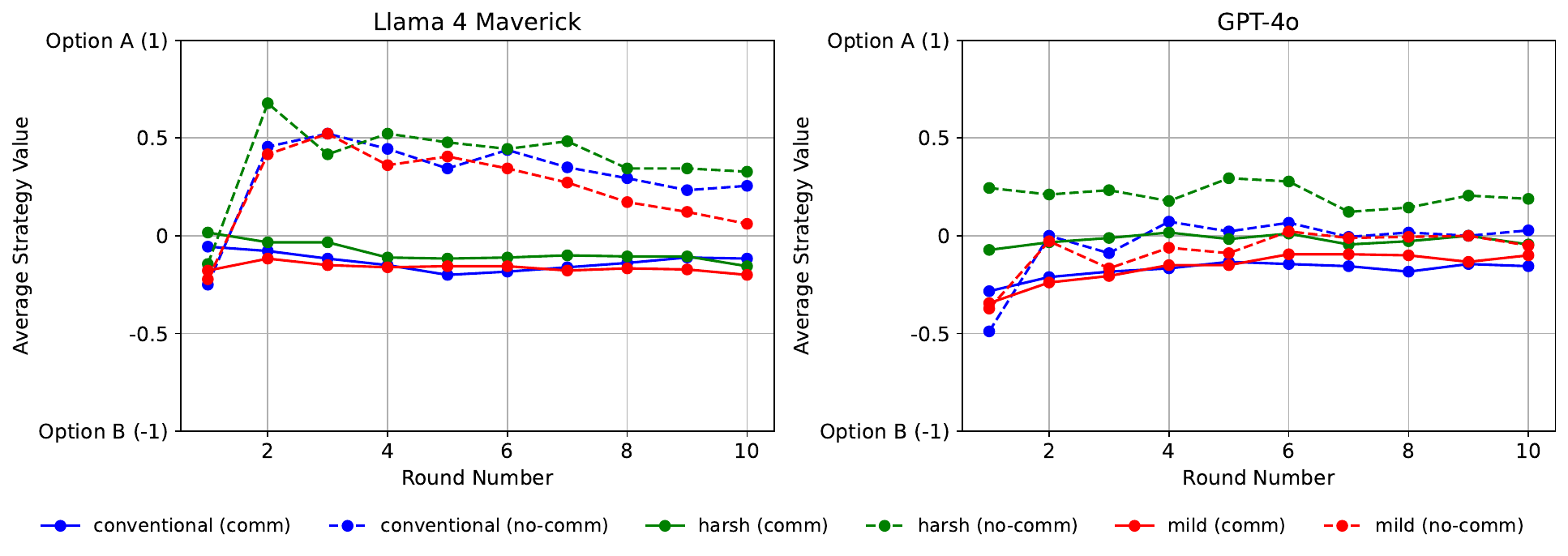}
    \vspace{-7mm}
    \caption{Strategy evolution over 10 rounds in the repeated Prisoner's Dilemma game, comparing LLMs with and without communication. Strategy values range from $+1$ (pure defection, Option A) to $-1$ (pure cooperation, Option B), averaged across games by personality type and communication setting.}
    \label{fig:linear_plots_battle}
    \vspace{-5mm}
\end{figure*}

\subsection{Battle of Sexes}
\label{sub:battle_sexes_results}

Figure 3 pairs Figure~\ref{fig:results_prisoner}, for the Battle of Sexes game. As explained in Section~\ref{sec:methodology}, the Battle of Sexes penalizes coordination failure more than defection, making alignment of preferences the preferred choice. %The results appear much more diverse compared to the Prisoner's Dilemma.
The figure displays the total final rewards accumulated by both agents, serving as an indicator of overall coordination success. Higher bars reflect stronger alignment and more effective coordination between the agents. As the figure illustrates, the results vary substantially across different languages and models, much more than in Figure~\ref{fig:results_prisoner}.

Notably, for Vietnamese (bottom row), Llama 4 Maverick consistently achieves equal or higher rewards when communication is enabled, suggesting that communication may facilitate coordination in this setting.
In contrast, in English and Arabic (middle and top row), Llama 4 agents often perform slightly better without communication, especially in one-shot games, implying that communication might introduce noise or misalignment in those languages.
As previously, selfish personality pairings yield lower rewards across conditions.
However, the results exhibit overall instability, reflected in wide confidence intervals. This instability may be partly due to the reduced number of test runs, only one third compared to the Prisoner's Dilemma, which was tested under three distinct payoff settings. Due to the computational cost of running the full factorial design, we opted for a reduced number of runs for the Battle of Sexes game.
Further investigation is needed to fully understand the underlying causes of these disparities.

\begin{figure*}[t]
    \label{fig:results_battle}
    \makebox[\textwidth][c]{%
        \includegraphics[width=1\linewidth]{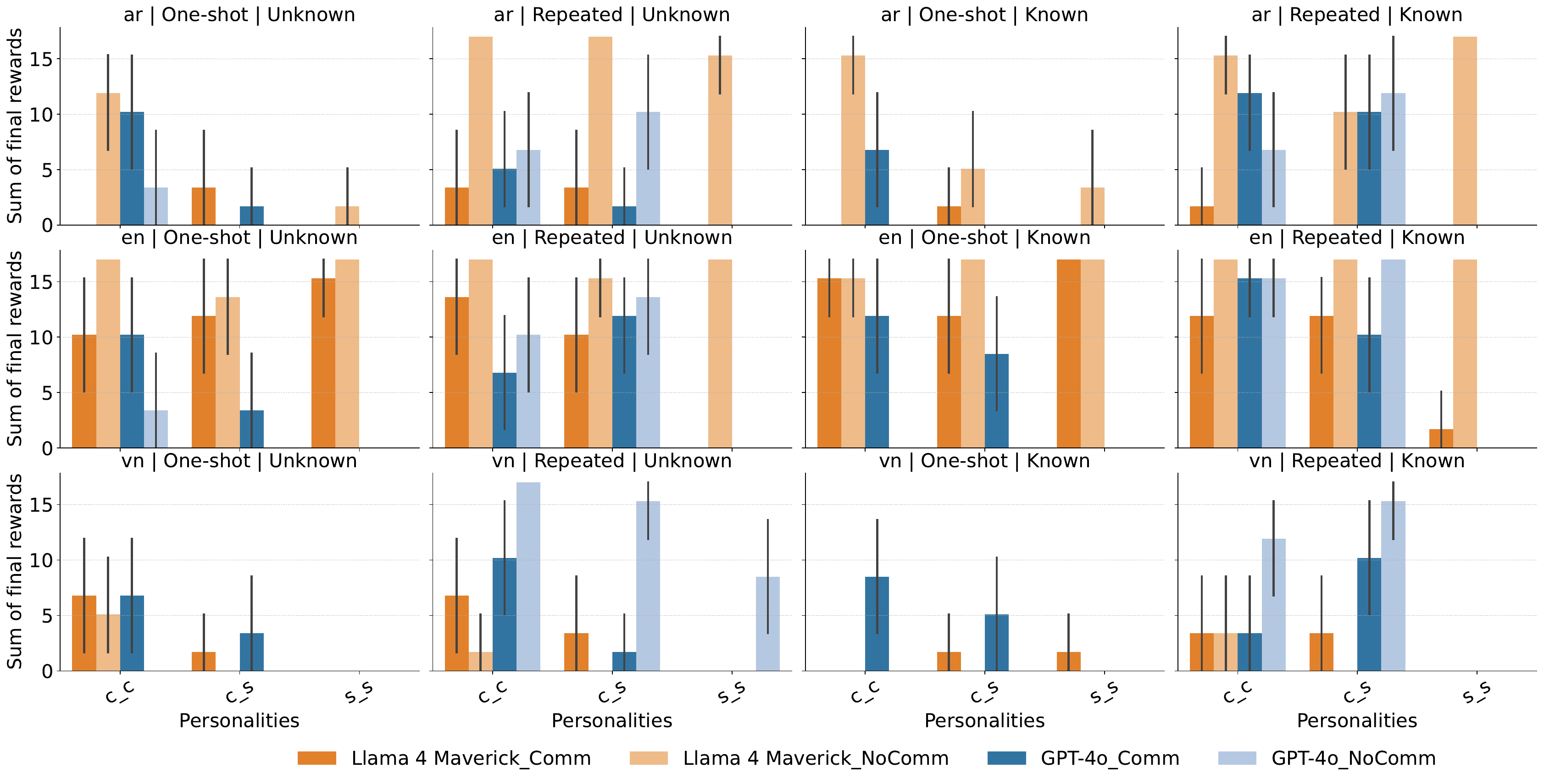}%
    }
        \vspace{-7mm}
    \caption{Impact of communication in the Battle of Sexes game across the considered LLMs. Each bar corresponds to the sum of the penalties obtained by both agents at the end of the game. Each plot corresponds to a combination of language | type of game (repeated/one-shot) | awareness of the number of rounds (known/unknown). 95\% confidence intervals are reported for each bar.}
        \vspace{-2mm}
\end{figure*}

Figure 4 shows the evolution of strategy coordination across rounds in the Battle of Sexes game. Unlike the Prisoner's Dilemma--where communication consistently enhances coordination--this figure highlights the opposite trend. GPT-4o, for instance, shows a marked improvement in coordination without communication, transitioning from nearly complete misalignment to moderate alignment -- as also observed in \cite{buscemi2025fairgame}. However, when communication is enabled, coordination only slightly increases, and overall remains lower than in the no-communication setting. For Llama 4 Maverick, a modest improvement is observed in the absence of communication, whereas enabling communication results in a slight decline in coordination over the course of the game.

\begin{figure}[t]
    \centering
    \includegraphics[width=0.7\linewidth]{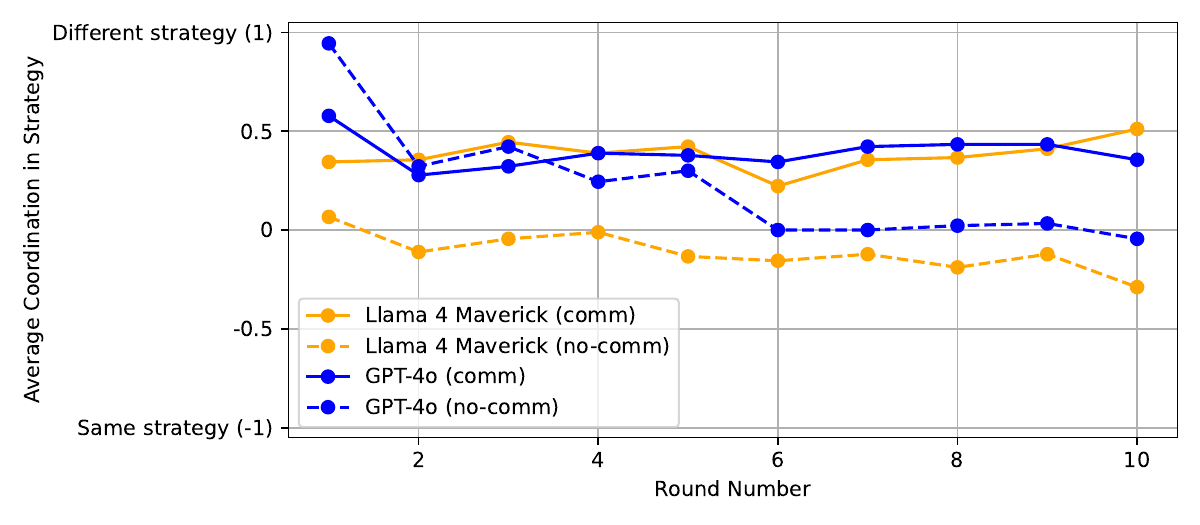}
        \vspace{-5mm}
    \caption{Average evolution of coordination in strategy choices across repeated rounds for all experiments of the Battle of Sexes games, for each LLM. Solid or dashed compare when communication is enabled and when it is disabled. Values represent alignment in option selection in each round, to achieve coordination. The value 1 corresponds to a mismatch in strategies (one selects Option A, the other Option B), reflecting coordination failure or defective behavior, while -1 indicates alignment in choices (successful coordination or cooperative behavior). 
    }
    \label{fig:linear_plots_battle_sexes}
        \vspace{-5mm}
\end{figure}

% \subsection{Scoring system}
% \label{sub:metrics}

% In our previous work, we introduced a set of evaluation metrics designed to quantify key behavioral characteristics of single LLMs and to map their strategic tendencies. These include measures of internal variability, cross-language inconsistency, sensitivity to payoff structures, and variability over rounds.
% These metrics are directly relevant to the present analysis, as they capture dimensions of behavior,such as consistency, adaptability, and linguistic robustness,that are critical for interpreting LLM performance in strategic multi-agent interactions. For this reason, we adopt them as core evaluation tools in the current study.
% Further details and formal definitions can be found in the cited paper.

\subsection{Analysis of the communication}
\label{sub:analysing}

We now examine in details how LLM agents adapt their communicative behavior, specifically message length, based on game type (repeated vs. one-shot), language, personality, and whether the number of rounds is known in advance. 

Table \ref{tab:merged_split_games} shows average message length per round across both games, highlighting the impact of knowing the interaction horizon. As expected, repeated games elicit longer messages than one-shot ones, reflecting more strategic or collaborative communication.
For Llama 4 Maverick, message length generally increases when the number of rounds is known, across all languages and games. This suggests anticipatory planning: knowing the interaction's duration, the agent invests more in signaling intent and fostering cooperation. For example, in the Prisoner's Dilemma, Arabic messages rise from 561.8 to 596.5 characters; in English, from 651.1 to 683.3. A similar pattern holds in the Battle of Sexes for Arabic and English, though Vietnamese exhibits a slight decrease.
Conversely, GPT-4o often shortens its messages when rounds are known, seen in Arabic (417.9 to 379.5) and Vietnamese (582.4 to 541.3) repeated games, suggesting a more execution-focused approach. Rather than negotiating, it may adopt more direct strategies once temporal limits are established.
Finally, Vietnamese messages are consistently longer than those in Arabic and English, likely due to linguistic verbosity rather than model behavior; similarly, Arabic may be shorter due to harakats being omitted in writing. These observations underline the importance of accounting for language-specific traits when interpreting message length as a proxy for strategy.

\begin{table}[h]
\centering
\scriptsize
\makebox[\linewidth]{
\begin{tabular}{|>{\centering}p{1.2cm}|>{\centering}p{1.3cm}|>{\centering}p{2.2cm}|>{\centering}p{2.2cm}|>{\centering}p{2.2cm}|>{\centering\arraybackslash}p{2.2cm}|}
\hline
\textbf{Lang} & \textbf{Rounds Known} & \textbf{PD Rep  (L4M/G4o)} & \textbf{PD OS (L4M/G4o)}  & \textbf{BS Rep (L4M/G4o)}  & \textbf{BS OS (L4M/G4o)} \\
\hline
ar & False & 561.8 / 417.9 & 273.3 / 212.8 & 575.15 / 405.57 & 312.77 / 213.50 \\
ar & True  & 596.5 / 379.5 & 306.2 / 193.6 & 611.02 / 366.94 & 342.70 / 205.10 \\
en & False & 651.1 / 520.5 & 340.5 / 273.7 & 667.20 / 526.73 & 415.27 / 284.40 \\
en & True  & 683.3 / 496.3 & 349.3 / 260.6 & 684.80 / 500.16 & 418.70 / 292.73 \\
vn & False & 722.8 / 582.4 & 352.0 / 310.2 & 709.68 / 584.60 & 398.80 / 315.77 \\
vn & True  & 704.1 / 541.3 & 314.9 / 286.5 & 704.17 / 538.31 & 386.23 / 329.03 \\
\hline
\end{tabular}}
\vspace{1mm}
\caption{Mean message length by language, rounds knowledge, game (Prisoner's Dilemma = PD; Battle of Sexes = BS) and game type (Repeated = Rep; One-shot = OS), comparing LLMs (Llama 4 Maverick = L4M, GPT-4o = G4o).}
\label{tab:merged_split_games}
\vspace{-8mm}
\end{table}

Beyond rounds knowledge and language, Table 3 reports average message lengths by personality pairing and game type, across both games and models, showing how social dynamics influence communication.
In the Prisoner's Dilemma, personality differences produce clear gradients. For Llama 4 Maverick, repeated messages drop from 670.2 (c\_c) to 627.0 (s\_s), and one-shot from 361.4 to 278.3, indicating that cooperative dyads engage in richer communication, likely to build trust or signal reciprocity. GPT-4o mirrors this pattern, though less strongly.
In contrast, the Battle of Sexes shows a weaker personality effect. Message lengths are more stable across pairs, especially in repeated games (e.g., Llama: 677.8 for c\_c vs. 653.9 for s\_s). Here, the coordination imperative outweighs prosocial signaling, and agents may focus more on alignment than persuasion.
Overall, these findings show that game structure and social roles shape LLM communication differently. The Prisoner's Dilemma amplifies the effects of personality and planning, whereas the Battle of Sexes prompts more uniform, coordination-oriented dialogue. Llama 4 Maverick tends to respond to strategic uncertainty with more adaptive verbosity, while GPT-4o favors a more concise, execution-driven style.

\begin{table}[t]
\centering
\scriptsize
\makebox[\linewidth]{
\begin{tabular}{|>{\centering}p{2cm}|>{\centering}p{2.2cm}|>{\centering}p{2.2cm}|>{\centering}p{2.2cm}|>{\centering\arraybackslash}p{2.2cm}|}
\hline
\textbf{Personality} & \textbf{PD Rep  (L4M/G4o)} & \textbf{PD OS (L4M/G4o)}  & \textbf{BS Rep (L4M/G4o)}  & \textbf{BS OS (L4M/G4o)}  \\
\hline
c\_c & 670.20 / 454.00 & 361.38 / 296.44 & 677.83 / 465.45 & 399.37 / 309.87 \\
c\_s & 662.50 / 512.50 & 328.39 / 269.25 & 654.91 / 492.23 & 373.35 / 266.35 \\
s\_s & 627.00 / 502.60 & 278.28 / 202.95 & 653.85 / 525.94 & 338.95 / 257.78 \\
\hline
\end{tabular}}
\vspace{1mm}
\caption{Mean message length by personality pair, game (Prisoner's Dilemma = PD; Battle of Sexes = BS) and game type (Repeated = Rep., One-shot = OS), comparing LLMs (Llama 4 Maverick = L4M, GPT-4o = G4o).}
\label{tab:merged_personality_lengths}
\vspace{-2mm}
\end{table}

Table 4 presents the average message length per round across different Prisoner's Dilemma payoff configurations (conventional, harsh, and mild) considering both LLMs and game types. In this case, no consistent pattern emerges linking the game's difficulty to the length of the exchanged messages.

\begin{table}[t]
\centering
\scriptsize
\label{tab:prisoner_type_game}
\begin{tabular}{|c|ccc|ccc|}
\hline
\multirow{2}{*}{\textbf{LLM}} & \multicolumn{3}{c|}{\textbf{Repeated}} & \multicolumn{3}{c|}{\textbf{One-shot}} \\
\cline{2-7}
 & \textbf{Mild} & \textbf{Conventional} & \textbf{Harsh} & \textbf{Mild} & \textbf{Conventional} & \textbf{Harsh}  \\
\hline
Llama 4 Maverick & 533.42 & 573.58 & 557.02 & 375.72 & 361.04 & 395.17  \\
\hline
GPT-4o           & 474.86 & 476.26 & 486.34 & 215.92 & 215.82 & 226.64 \\
\hline
\end{tabular}
\label{tab:prisoner_configuration}
\vspace{1mm}
\caption{Prisoner's Dilemma – Mean message length by LLM, communicative style, and game type.}
\vspace{-2mm}
\end{table}

\begin{figure}[t]
    \centering
    \includegraphics[width=0.7\linewidth]{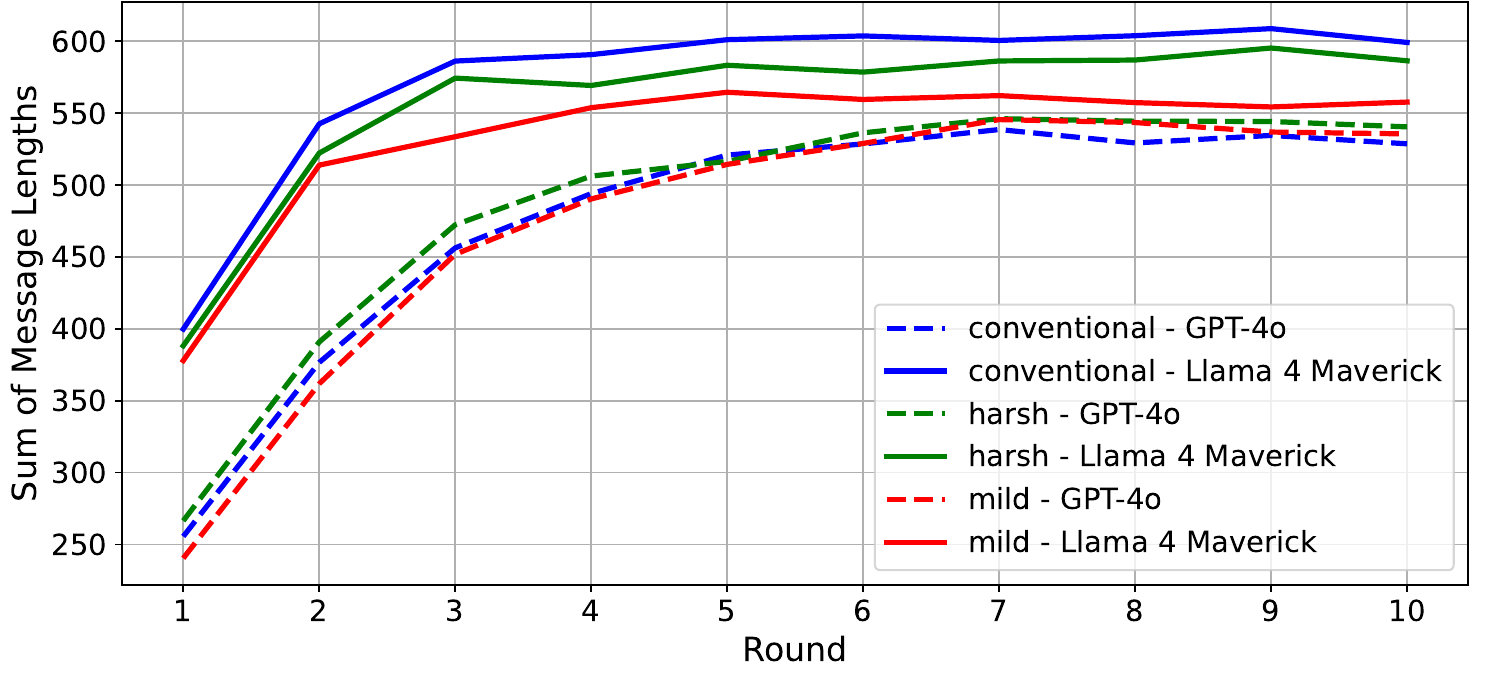}
     \vspace{-5mm}
    \caption{Total message length, defined as number of characters, per round in repeated Prisoner's Dilemma games, across payoff types and LLMs.}
    \label{fig:prisoner_messages}
    \vspace{-5mm}
\end{figure}

Figure \ref{fig:prisoner_messages} shows total message length per round (sum of both agents) in repeated Prisoner's Dilemma games across three payoff configurations.
Message length rises sharply from round 1 to 2, indicating an early phase of strategic alignment or signaling, followed by a plateau from round 3 onward as agents settle into stable behaviors.
Llama 4 Maverick maintains more variable communication, especially in conventional and harsh settings, with fluctuations persisting through later rounds. GPT-4o, by contrast, stabilizes more quickly, suggesting a front-loaded or more rigid communicative strategy.
Differences between payoff configurations are less evident early on, particularly for Llama, but increase over time, suggesting that dilemma severity shapes the communication.

Table 5 lists the five most frequent non-generic words used by LLMs in Arabic, Vietnamese, and English across all configurations of the Prisoner's Dilemma and the Battle of Sexes, both repeated and one-shot. Words were extracted using the Python library spaCy \cite{honnibal2020spacy}, excluding stopwords and game-specific terms such as "Option A", "Option B", "choose", and "round". Although the experiments involving Arabic use the Arabic script for both prompts and LLM outputs, the words in the table are transliterated into the Latin alphabet to enhance readability for non-Arabic readers and ensure cross-linguistic comparison.

In the Prisoner's Dilemma, frequent terms such as "penalty", "trust", "cooperation", and "outcome" highlight a focus on risk, reciprocity, and strategic foresight--key aspects of social dilemmas. Future-oriented verbs like saʾakhtār ("I will choose") further emphasize this planning dimension.
In contrast, the Battle of Sexes features vocabulary related to reward, preference, and emotional or evaluative terms such as "good", "think", and "appreciate", which align with the game's emphasis on coordination and mutual satisfaction. The recurrence of inclusive expressions like “cùng" (“together”) in Vietnamese across both games suggests a consistent drive toward joint decision-making.
These lexical patterns underscore how LLMs adapt their language to the underlying incentive structures of different games, revealing human-like sensitivity to context and strategy.

\vspace{-5mm}
\begin{table}[h]
\centering
\scriptsize
\label{tab:merged_word_freq}
\makebox[\linewidth]{%
\begin{tabular}{>{\centering}p{2.2cm} >{\centering}p{2cm} >{\centering}p{1.8cm} | >{\centering}p{2.2cm} >{\centering}p{2cm} >{\centering\arraybackslash}p{1.8cm}}
\toprule
\multicolumn{3}{c|}{\textbf{Prisoner's Dilemma}} & \multicolumn{3}{c}{\textbf{Battle of Sexes}} \\
\textbf{Arabic (transl.)} & \textbf{Vietnamese} & \textbf{English} & \textbf{Arabic (transl.)} & \textbf{Vietnamese} & \textbf{English} \\
\midrule
ʿuqūba (penalty) & phạt (penalty) & penalty & mukāfa'a (reward) & chọn (choose) & time\\
afḍal (better) & cùng (together) & trust &  al-akhirā (the last) & cùng (together) &  preference\\
al-taʿāwun (cooperation) & hợp (match) & let & jayyida (good) & nhận (receive) &  think \\
natīja (result) & cả (all/both) & good &  ikhtiyār (selection) & thưởng (reward) &  final \\
saʾakhtār (I will choose) & hai (two) & outcome &ʾaʿtaqid (I think) & phần (part) & appreciate \\
\bottomrule
\end{tabular}%
}
\vspace{1mm}
\caption{Frequent multilingual words used by LLMs in Prisoner's Dilemma and Battle of Sexes games (with translations)}
\vspace{-10mm}
\end{table}

\section{Conclusion and Future Directions}
\label{sec:conclusion_future}

This study systematically investigates how communication influences the strategic behavior of LLM-based agents in multi-agent environments. By extending the FAIRGAME framework with a communication layer, we analyzed the effects of message exchange in two canonical game-theoretic scenarios, the Prisoner's Dilemma and the Battle of Sexes, across multiple languages, personality profiles, and game configurations. Comparing Llama 4 Maverick and GPT-4o, we found that communication substantially alters agent behavior: in the Prisoner's Dilemma, a social dilemma involving trust and self-interest, communication fosters cooperation and improves outcomes; by contrast, in the Battle of Sexes, a coordination game, communication unexpectedly reduces the level of alignment between agents.

Our analysis reveals that communicative behavior is deeply context-sensitive. Message length, lexical choices, and strategic alignment vary with game type, personality dynamics, and language. These patterns underscore the dual role of language in LLM interactions: as a vehicle for coordination and as a potential vector for emergent biases.
This work lays the groundwork for a broader research agenda at the intersection of language, strategy, and AI coordination. Future investigations should extend to a wider set of languages, especially low-resource or typologically diverse ones, to assess the generalizability and fairness of communicative strategies. Clusters of related languages may illuminate whether linguistic similarity drives behavioral convergence, or whether model-specific traits dominate.
Deeper semantic analysis of agent communication remains a critical next step. While message length and frequency offer useful signals, uncovering strategic intent and negotiation dynamics will require more sophisticated tools. Advances in LLM-based evaluators and conversation modeling could enable scalable, nuanced interpretation of agent interactions, though such methods must also confront issues like evaluator bias.
Ultimately, understanding how LLMs reason, negotiate, and cooperate through language in complex social settings is key to building robust and trustworthy multi-agent systems. This study marks a step in that direction, emphasizing that communication is not an auxiliary feature but a central mechanism shaping AI behavior.
These findings inform the design of safer and more interpretable multi-agent AI systems, especially in applications where strategic alignment and fairness are critical \cite{capraro2024language,buscemi2025fairgame,han2021or}.

%%%%%%%%%%%%%%%%%%%%%%%%%%%%%%%%%%%%%%%%%%%%%%%%%%%%%%%%%%%%%%%%%%%%%%%%

%%% Use this environment to include acknowledgements (optional).
%%% This will be omitted in doubleblind mode.

% \begin{ack}
% D.P. is supported by the European Union through the ERC INSPIRE grant (project number 101076926). Views and opinions expressed are hfwever those of the authors only and do not necessarily reflect those of the European Union or the European Research Council Executive Agency. %Neither the European Union nor the European Research Council Executive Agency can be held responsible for them. 
% T.A.H. is supported by EPSRC (grant EP/Y00857X/1).

% FAIRGAME is open-source and accessible on Github  \cite{githubFairgame}, with examples of configuration files and templates.
% \end{ack}

%%%%%%%%%%%%%%%%%%%%%%%%%%%%%%%%%%%%%%%%%%%%%%%%%%%%%%%%%%%%%%%%%%%%%%%%

%%% Use this command to include your bibliography file.

\end{document}